# Integrating Evidence into the Design of Explainable Artificial Intelligence and Decision Support Systems: A Means-End Framework for End-users in Construction

Peter E.D. Love [a], Jane Matthews [b], Weili Fang [c], & Hadi Mahamivanan [d]

[a] School of Civil & Mechanical Engineering, Curtin University, GPO Box U1987, Perth, Western Australia, 6845, Australia, Email: p.love@curtin.edu.au

[b] School of Architecture & Built Environment, Deakin University, Waterfront Campus, Geelong, VIC3227, Australia, Email: jane.matthews@deakin.edu.au

[c] School Civil Engineering & Hydraulic Engineering, Huazhong University of Science & Technology, Wuhan, 430074, China, Email: weili_Fang@hust.edu.au

[d] School of Architecture & Built Environment, Deakin University, Waterfront Campus, Geelong, VIC3227, Australia, Email: s222342928@deakin.edu.au

# Integrating Evidence into the Design of Explainable Artificial Intelligence and Decision Support Systems: A Means-End Framework for End-users in Construction

**Abstract:** Explainable Artificial Intelligence (XAI) aims to make the reasoning processes of AI models transparent and interpretable, especially in complex decision-making environments. In construction, where AI-based decision support systems (DSSs) are increasingly deployed, limited attention has been given to incorporating *evidence* to support the reliability and accountability of their outputs. Without embedding evidence into the design of XAI and DSSs, the validity of explanations and the trustworthiness of system recommendations remain open to question. To address this challenge, this paper presents a theoretical, evidence-based *means–end* framework developed through a narrative review. The framework provides an epistemic foundation for designing XAI-enabled DSSs that can produce *meaningful human explanations* (MHEs) aligned with end-users' knowledge needs and decision-making contexts. It emphasizes evaluating the *strength*, *value*, and *utility* of different types of evidence used to support explanations generated by AI systems. While the framework is designed with a focus on construction professionals as primary end-users, its relevance extends to other stakeholders, such as developers, regulators, and project managers, whose epistemic goals may differ. By explicitly integrating evidence into the development of XAI and DSSs, the framework enhances transparency, fosters trust, and improves the effectiveness of decision-making. Given its theoretical foundation and broad applicability, the framework can inform the design of evidence-aware XAI systems across various industrial and engineering domains.

## 1. Introduction

Artificial intelligence (AI) is having a profound influence on all aspects of work, and with advances being made at an accelerated rate, a myriad of opportunities (e.g., task automation, predictive analysis, and real-time decision-making) and challenges (e.g., job displacement, data bias, and algorithmic fairness and data privacy and security) manifest [1]. Indeed, AI is a transformative technology that is reshaping the workplaces of various industrial sectors, including finance, healthcare, manufacturing, and transportation [1-3]. Still, within construction, the context for our paper, the rate of AI adoption has been cautious despite its potential benefits and widespread uptake in other sectors [1], [2]. A key barrier to AI adoption in construction is the limited understanding of *how* AI systems reach their conclusions, which fuels mistrust and resistance among practitioners. Construction projects are characterised by high uncertainty, fragmented teams, and context-specific decisions, making transparency and accountability crucial for effective decision-making.

The hesitancy in construction organizations' adoption of AI is not rooted in reluctance, but rather in a lack of understanding about what benefits AI can offer, *why* they matter, and *how* and *when* these can be realized without disrupting ongoing operations [2]. While it is also acknowledged that AI can improve decision-making by quickly processing large amounts of information, some construction organizations' managers are mindful of the risk of *automation bias* [2], [3], [4]. In such instances, people may over-rely on automated systems and technologies, prematurely accepting AI-generated recommendations without adequate scrutiny, often leading to errors or suboptimal outcomes [3], [5]. This bias occurs when users place excessive trust in automated tools, such as AI algorithms and decision support systems[1]

[1] In this paper a DSS is a computer-based system leverages AI technologies to assist decision-makers in analyzing data, evaluating scenarios, and making informed decisions. A DSS that incorporates AI results in automating tasks, learning from data, and providing predictive and prescriptive insights. Business intelligence is a subset of DSS. Here decision intelligence seeks aims to help automate decision-making using AI. The goal of decision intelligence is to design, model, align, execute, monitor, and tune decision models and processes. Off-the-shelf examples of DSSs available in construction are Autodesk's Construction IQ, Kwant.ai and Trackunit IrisX to name a few.

(DSSs), and undervalue their own expertise or contextual knowledge, even when the automated output may be unreliable or inappropriate [3].

Explainable Artificial Intelligence (XAI) offers a means to mitigate these concerns by making AI decisions more transparent, interpretable, and trustworthy. In doing so, XAI can support critical engagement with AI-generated outputs, reduce undue dependence on automation, and foster more responsible, context-aware decision-making. Accordingly, this paper focuses on the potential of XAI to facilitate the effective and accountable deployment of AI-driven DSSs in construction, ultimately enhancing the effectiveness of human–machine collaboration throughout all phases of a project's life cycle.

Many AI-based DSSs developed for construction, whether bespoke or off-the-shelf, operate as 'black boxes', producing recommendations or insights without offering explanations of their internal logic or decision-making processes [6], [7]. This opacity is particularly problematic with generative AI models (e.g., Generative Adversarial Networks, Variational Autoencoders, Reinforcement Learning, and Transformer-based models), which are increasingly used to underpin DSSs by generating data-driven content and scenarios for a range of tasks such as scheduling, cost estimation, identification of safety risks, and assessing quality [6–13]. Without explainability, users may lack the confidence or context needed to critically evaluate these outputs. XAI, therefore, plays a crucial role by making AI system behavior intelligible through explanations, allowing users to understand, scrutinise, and ultimately trust the decisions generated by AI-based DSSs [6]. Thus, calibrating user trust to the true capabilities of AI through transparent and context-aware explanations is essential to prevent over-reliance or undue skepticism.

An in-depth review of XAI in construction, specifically examining its core precepts (i.e., explainability and interpretability), its relationship to different types of AI models (e.g., transparent and opaque), and levels of post-hoc explainability (e.g., model-agnostic and model-specific), is provided by Love *et al*. [6]. However, this review adopts a predominantly technological and application perspective, offering limited discussion of user-centred design practices grounded in empirical 'evidence' and systematic evaluation techniques. These aspects are increasingly recognised as critical to the development of effective and trustworthy decision support systems [2], [11–13]. Similar gaps are evident in several XAI reviews across other domains, where technical concerns often overshadow the practicalities of how explanations are designed, delivered, and evaluated from the end-user's perspective [14–22].

As XAI continues to advance, growing concerns have emerged regarding the quality of evidence used in the design of DSSs. Many systems are built on flawed or low-quality evidence, influenced by confounding variables, biased training and test data, or overfitting, factors that can significantly distort interpretability and reliability [9], [10], [23], [24–26]. A striking example is provided by DeGrave *et al*. [27], who showed that deep learning models developed to detect COVID-19 from chest radiographs relied on confounding features unrelated to pathology. As a result, the models appeared highly accurate but failed when deployed in new clinical settings, a failure that highlights the dangers of relying solely on statistical associations in the presence of confounders. This case exemplifies the broader risk of drawing causal inferences from associative evidence when confounding variables are not adequately addressed [28].

Association studies alone are insufficient to produce reliable explanations of outcomes generated by AI models [28]. As Durán [26] persuasively argues, no single scientific explanation, based on statistical associations, can fully capture or account for empirical reality, implying that explanatory pluralism is essential for informed decision-making. In the context of construction, where XAI is gaining momentum, the strength of evidence underpinning many models remains weak, often grounded in expert opinion or limited case study demonstrations. Consequently, the validity and generalizability of the explanations derived from these AI systems are open to question [29–47].

With construction organizations at the cusp of the AI adoption technology cycle and XAI beginning to receive attention, albeit at a slow pace compared to other fields (e.g., law and medicine), an opportunity exists to take heed of contemporary thinking and learn from their insights and experiences [26-28], [48]. Thus, if construction organizations are to take advantage of the benefits of AI, DSSs must be interpretable and transparent *ab initio* in their design so that trustworthy and ethical solutions can be generated [2], [6], [49].

Using a narrative review, this paper addresses the research question: *How can evidence effectively inform the development of XAI systems and DSSs to ensure the epistemic needs of end-users are met?* Three sub-questions guided our inquiry. First, we examine *the role of evidence in XAI system design*, highlighting how different types of evidence, such as those from association and mechanism studies, support the reliability and interpretability of AI outputs, and advocate for a hierarchy of evidence to guide system development. In this context, *evidence* refers to credible, traceable, and justifiable information or reasoning that underpins the reliability, transparency, and practical value of AI explanations. It is positioned as the epistemic foundation that validates, or warrants, the outputs and recommendations of AI models and DSS, thereby enhancing user trust, interpretability, and decision quality.

Second, we examine *how XAI can deliver explanations that meet users' epistemic needs*, introducing the concept of Meaningful Human Explanations (MHEs) —those that are intelligible, relevant, and actionable, enabling users to understand both why and how AI-generated outputs are produced [23], [24]. Finally, we address *how construction organizations can engage with XAI despite not being system developers* by demonstrating how end-users can understand AI-driven decisions, articulate epistemic requirements to vendors and researchers, and critically assess DSSs to ensure alignment with their practical needs.

Our paper is structured as follows. It begins by explaining and justifying the research approach adopted (Section 2). The concept of XAI is briefly examined to establish the paper's setting. However, a detailed synthesis within the context of construction can be found in Love *et al*. [2] [6] and Naser [38] (Section 3). A conceptual and theory-building means-end framework is then introduced, based on the emergent developments of XAI from an extensive range of literature (e.g., computer science, philosophy, and medicine), to provide a structure for effectively utilizing evidence in designing DSSs. Hence, ensuring that outputs are explainable and enhancing end-user decision-making (Section 4). The introduction of the means-end approach provides a segue to examine the nature of evidence required to support the design and development of XAI and DSSs (Section 5). Then, the implications of a means-end approach to support the design of DSSs in construction are discussed, with the paper's limitations and avenues for future research being identified next (Section 6) before presenting the paper's conclusions and contributions (Section 7).

## 2.0 Research Approach

While the computer science and information systems literature, for example, are embracing the need for evidence to be incorporated into the design of more effective DSSs and provide understandable explanations of AI-generated outputs, the construction and engineering management (CEM) literature remains silent on the matter [9], [10]. Given the limited body of work on XAI in the CEM literature and the lack of evidence supporting explanations and DSSs design, a systematic review is not an appropriate approach to address this paper's research questions [50-57]. However, a narrative review is better suited for under-researched areas and where new insights or ways of thinking are required [59]. Narrative reviews are suitable for synthesizing a broad body of research on complex issues, where a detailed and nuanced description and interpretation are necessary to advance new ideas, as presented in this paper [60].

While there are no consensus reporting guidelines for conducting narrative reviews, as there are for systematic reviews, the research process adopted is outlined in Figure 1 [58-60]. The initial search focused on examining the content of XAI reviews and their applications to gain an understanding of developments and research gaps. The research question provided the guardrail for developing the search strategy using Scopus™, Google Scholar, Web of Science™, the Cornell University open access repository of preprints and postprints (i.e., arXiv), and the table of contents searches in journals. The search process was iterative, with additional keywords added as part of a trial-and-error approach to selecting appropriate articles (including those from mainstream computer science conferences) for inclusion in the review. Numerous books from established publishers are also utilized to reinforce and provide supplementary evidence to support our line of inquiry, which covers a range of fields and journals (Figure 1):

- *Construction engineering and management* (e.g., Advances in Engineering Software, Automation in Construction, AIC, ASCE Journal of Management in Engineering, ASCE

Journal of Computing in Civil Engineering, ASCE CIE, Reliability and Engineering Safety, RESS, and IEEE Transactions on Engineering Management, IEEE TEM);

- *Computer science* (e.g., Artificial Intelligence, AI, Advanced Engineering Informatics, AEI, Artificial Intelligence Review, Information Fusion, Computers in Industry, Nature Machine Intelligence, SIAM Journal on Computing, International Journal of Human-Computer Studies, and the Proceedings of the ACM on Human-Computer Interaction)
- *Philosophy* (e.g., International Studies in the Philosophy of Science, Phil. of Science, Studies in History and Philosophy of Science, Synthese, and the Stanford Encyclopedia of Philosophy) and
- *Medicine* (e.g., BMC Public Health, Computers in Biology and Medicine, European Journal of Clinical Investigation, Journal of Clinical Epidemiology, and PLoS Medicine).

The authors of this paper comprise acknowledged experts from various fields, including civil engineering, computer science, digital engineering, information systems, and management, who were subsequently drawn upon to determine whether an article should be included in the review based on the research question identified in Figure 1.

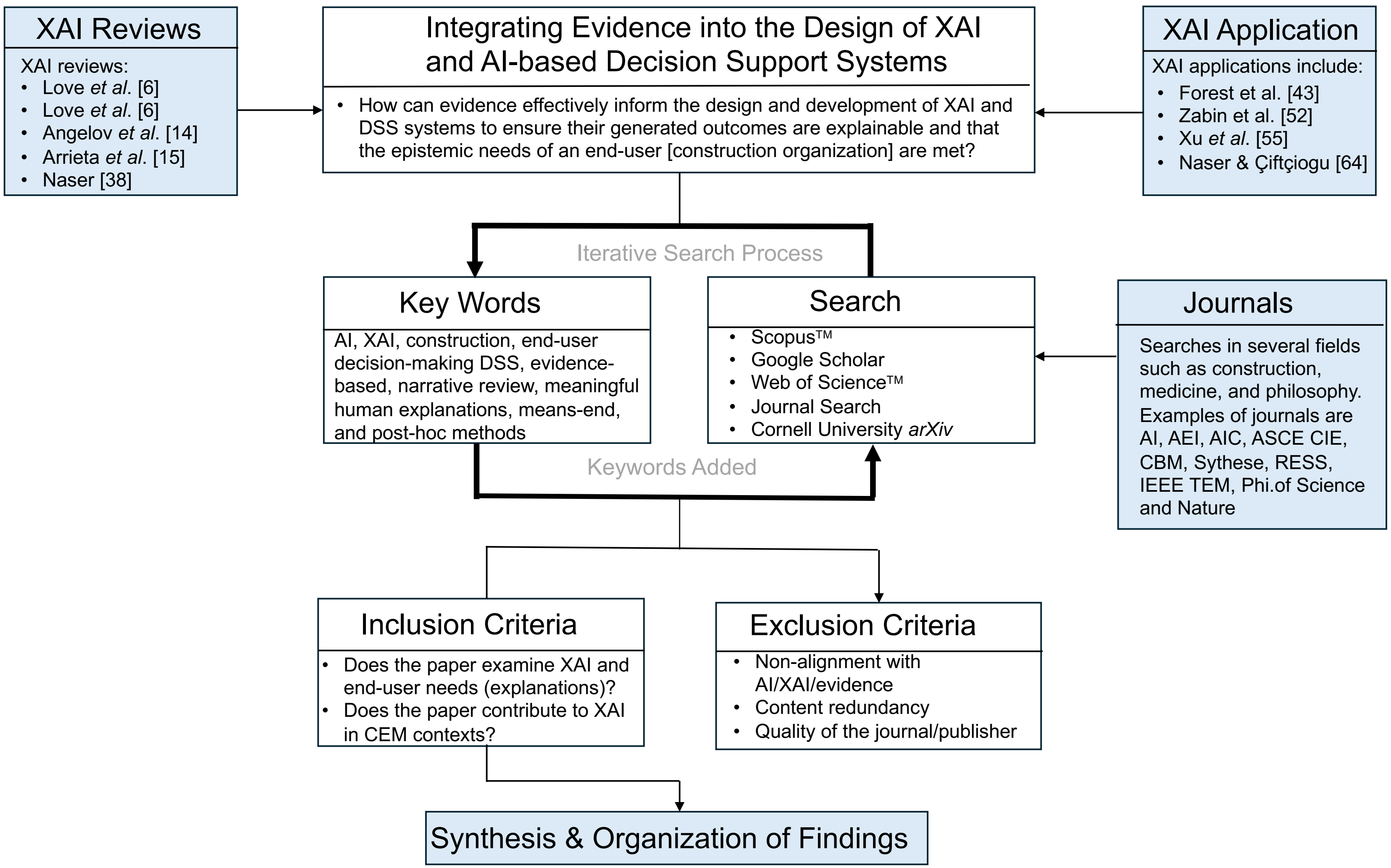

Figure 1. Review process

The authors were reflexive during the selection of articles, striving to maintain a neutral stance when deciding on their inclusion in the review and the development of the proposed means-end framework. Furthermore, including multiple authors with diverse backgrounds helps to bring balance to the synthesis of the literature.

## 3.0 Explainable Artificial Intelligence

The concept of XAI is concerned with developing techniques to understand the 'black box' workings of AI models [6], [8], [10], [11-23]. Thus, XAI aims to transform AI models (including their subset of machine learning) from untrustworthy 'black boxes' into transparent, understandable, and trustworthy tools. An array of XAI instruments has been developed to (partly) eliminate the perceived opacity of AI models [15]. As opaque AI models lack inherent interpretability, post-hoc methods generate explanations (e.g., feature importance scores, saliency maps, and SHapley Additive exPlanations, SHAP). Two of the most well-known and widely used XAI instruments in construction are [6], [14], [15], [61], [62-65]:

1. *Local interpretable model-agnostic explanations* (LIME): Explanations are *local,* explaining predictions for only a specific part of the data rather than the model's global behavior. For example, a model might be approximated locally by a match with an intuitive linear function (e.g., regression algorithms such as Lasso and Ridge), rendering it interpretable as its behavior can be understood [24], [66]. As LIME is model agnostic, it can provide outputs for several AI methods [67]. However, Garreau and von Luxburg [68] point out that data perturbations and poor choices in model parameters can result in LIME's missing important features, making the generated explanations subject to variability.

Hypothetically, for example, a construction manager uses a DSS to predict the likelihood of a project delay. The model predicts a high risk (82%) of a delay occurring. The top *five* factors influencing this prediction are:

(i) Subcontractor Reliability Score = Low (↓), projects with low subcontractor scores had 65% high delay rates

(ii) Design Finalization – Delayed (↓), delayed designs contributed to delays in 72% of past cases

(iii) Project Complexity Index = High (↑), high complexity added uncertainty and coordination issues

(iv) Rainfall Forecast = Above average (↑), weather delays were 30% more likely under similar forecasts

(v) Site Readiness Score = Medium (↔), medium scores moderately contributed to delay risks

As a result of these outputs, the construction manager may consider reassessing the subcontractor's capability, expediting the design's sign-off, or mitigating weather impacts by adjusting the project's schedule.

Numerous studies in construction have utilized LIME explanations. For instance, Leuthe *et al*. [32] leverage XAI to inform end-users (i.e., property owners) about how AI models predict building energy consumption, using LIME to explain the impact of multiple features on their performance. In a similar vein, Thisovithan *et al*. [46] utilize machine learning models to predict the fundamental period of masonry infill in reinforced concrete framed structures using both LIME and SHAP to visualize and provide an understanding and explanation of why an RF was the ‘best fit’ model; and

2. *Counterfactual explanations* (CFE): Provide a link between what actually happened and what could have happened had an input to a model been changed in a particular way. These explanations are typically generated after a final machine learning model is produced, addressing the question: ‘What would need to be different for a better outcome?’ [40], [64]. A CFE characterizes an outcome by exploring alternative scenarios or events that did not occur but plausibly could have. In XAI, CFEs focus on the input data, evaluating how changes to one or more features affect the predicted output. Designing CFEs is non-trivial, as it requires understanding the model’s decision boundaries and adjusting inputs just enough to shift the outcome meaningfully [69]. Effective CFEs should be actionable, grounded in real-world evidence, and diverse, revealing trade-offs and causal implications across scenarios [69]. Because CFEs rely only on the input–output relationship, they can intuitively explain model behavior without requiring access to the model’s internal logic, thereby sidestepping the complexity of its mathematical formulation [24]. This makes CFEs particularly useful when users need to understand how to intervene to avoid undesirable outcomes, such as reducing the risk of a safety incident.

For example, a site supervisor using a DSS may find that, based on projected workload and site conditions, there is a 72% likelihood of a serious incident occurring on a high-rise project. A counterfactual scenario might suggest that reducing the proportion of inexperienced workers from 40% to below 20% and increasing scaffolding inspections from weekly to twice weekly would decrease the predicted risk to 28%. Thus, adjusting workforce composition and inspection protocols could significantly reduce the likelihood of an incident.

Several studies have adopted CFE methods in construction-related applications. For example, Zhan *et al*. [40] utilized CFEs to enhance the understanding of risk-based decision-making in managing ground settlement during tunnel construction. Meanwhile, Naser and Çiftçioglu [64] applied causal discovery techniques to assess the fire resistance of reinforced concrete columns, showing that integrating causal inference into ML models can significantly improve interpretability.

In essence, XAI seeks to design and develop DSSs capable of performing complex inferences while generating explanations, through natural language, visualizations, or feature-based methods, that enable users to understand, trust, and act upon AI outputs. Central to this objective is the provision of explanation mechanisms grounded in evidence. However, as previously discussed, current XAI approaches applied in the design and development of DSSs in construction often lack such evidentiary grounding. This absence undermines both the transparency and trustworthiness of the systems. Embedding evidence into the design of XAI is therefore critical to ensuring that DSS outputs are not only interpretable but also reliable and justifiable, thereby enabling stakeholders, including developers, regulators, business managers, and end-users, to make well-informed, defensible decisions [2].

While it is important to consider a broad range of stakeholders in the design of XAI and DSSs, this paper primarily focuses on meeting the epistemic goals of end-users within construction organizations. Nonetheless, the issues addressed are also relevant to other stakeholders, whose evidence needs and explanation requirements may vary according to their distinct roles, objectives, and responsibilities.

## 4.0 Means-End Epistemology

Explanations, in the context of AI, have traditionally been interpreted in two primary ways [71–78]: (1) as intrinsic to the reasoning process itself, and (2) in terms of their usage and functionality, aiming to make the model's reasoning and outputs understandable to users. In the construction domain, much of the prevailing AI research emphasizes high-level classification tasks, such as labeling, clustering, and pattern recognition, to facilitate the interpretation of DSSs and their outputs [78–88]. For example, Zhong *et al*. [83] apply deep learning techniques (e.g., bi-directional long short-term memory and conditional random fields) to automatically extract and classify procedural information from quality documents for compliance checking. Similarly, Wang *et al*. [85] use a combination of RF-SMOTE and SHAP to classify defect-related text in reports, thereby helping engineers understand the rationale behind a model's predictions.

However, while such approaches highlight key features and assign weights to inform decision-making, the relationship between input features and classification outputs is often indirect and unstable [89]. Minor changes in seemingly unrelated aspects of input data can lead to different rankings of feature importance, which in turn affect model outputs. Consequently, slight variations in initial conditions or data preprocessing can result in the construction of substantially different models [89], raising concerns about the consistency, transparency, and reliability of AI-generated explanations of DSSs in construction.

While classifications are a type of model output, indicating how input data has been assigned to predefined categories or labels, they are not explanations, as they do not reveal the reasoning or evidence behind a model's decisions [26]. A cursory examination of AI research in the CEM literature reveals that such outputs are often accepted at face value, with limited reliance on bona fide scientific evidence to explain or justify the results produced [24], [29–47], [78].

A case in point can be found in the work of Mostifi *et al*. [90], who use an ensemble 'voting' classifier, a machine learning technique, to classify and predict rework costs (i.e., corrections in construction from 2,527 non-conformance reports). However, prevailing evidence suggests that machine learning techniques are unable to accurately predict rework costs. Rework is not a risk but an uncertainty, which is probabilistically unmeasurable [91]. Non-conformances requiring rework typically arise due to human errors, which cannot be predicted as project environments, prone to volatility, uncertainty, complexity, and ambiguity (VUCA), juxtaposed with workplace conditions, influence their occurrence [92], [93]. Moreover, non-conformances are often scant in detail and incomplete, with causes being shoe-horned into artificially constructed categories that do not reflect the context and actualities of an event [79].

### *4.1 A Means-End Framework*

Regardless of how an explanation is interpreted, it is important to recognize that the concept itself is inherently malleable and context-dependent [78]. While the philosophical nuances of explanation are beyond the scope of this paper, they are explored in depth by Salmon [72], Woodward and Ross [94], and Sørmo *et al*. [95]. For the purposes of this study, we adopt Salmon's [72] epistemic conception of explanation, as supported by Buchholz [24], which frames explanation as fundamentally connected to human understanding and knowledge.

Under this conception, an explanation is considered successful if it enhances an individual's understanding of a phenomenon. It prioritizes the cognitive and informational dimensions of explanation, emphasizing its role in improving a person's epistemic state. By focusing on the 'why' or 'how' something occurs, this view enables individuals to make sense of events within a specific context and to integrate new information meaningfully with what they already know. Developing such explanations requires not only a robust body of evidence but also careful consideration of the means by which they are generated. These means, including the methods, processes, and reasoning applied, can vary depending on the epistemic ends, or the goals related to knowledge and understanding, that are being pursued.

To ensure explanations are both valid and useful, a means-end epistemology is needed, one that critically evaluates how evidence is used to achieve epistemic aims [24], [72]. Within this framework, epistemic normativity plays a central role, addressing fundamental questions about *what should be believed*, *how beliefs should be formed*, and *what counts as sound justification*. It provides the standards that guide our cognitive efforts, ensuring that reasoning processes and inquiry methods align with principles that foster knowledge, truth, and rational understanding.

A means-end epistemology's central normative criterion – the standard or principle that determines the epistemic strengths and shortcomings of a belief, reason, or method of inquiry – is explained by the principle of *instrumental rationality* [24], [96]. In this instance, individuals aim to select the most effective means to achieve their goals. Thus, to design and develop DSSs that are explainable and interpretable for end-users in construction, a means-end framework, as shown in Figure 2, is proposed.

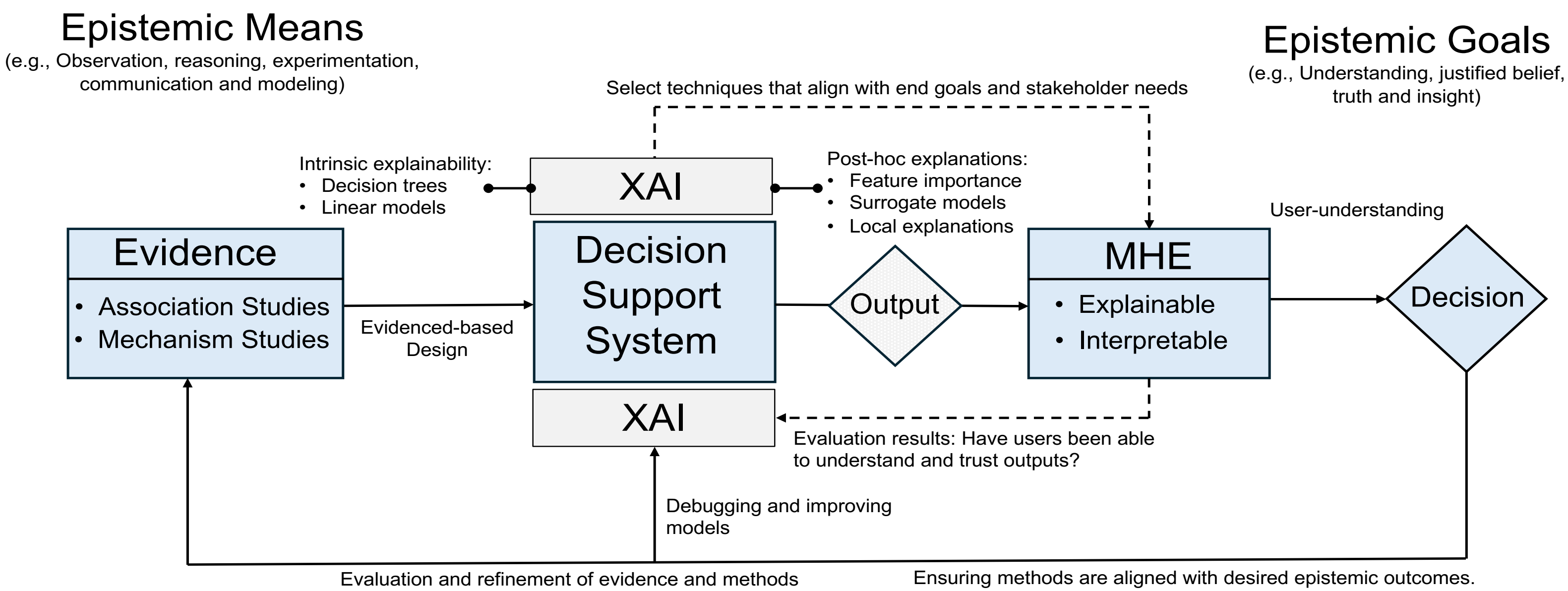

Examples of *how* evidence can inform XAI instruments:

- ***Guiding Feature Relevance and Model Structure***: Ensures input features and model architecture reflect domain-specific risks, causal relationships, and priorities. Historical safety reports and accident logs reveal that incidents are more frequent when inexperienced workers are present on-site and when overtime exceeds standard thresholds. Safety incident data shows strong links between overtime and incident rates, prompting the inclusion of 'overtime hours' and 'worker experience' as key model inputs. ***XAI Techniques*** – Domain-informed feature engineering, Feature importance ranking *via* SHAP and permutation methods) and causal inference methods (e.g., CausalNex)
- ***Shaping Explanation and Delivery*****:** Adapts explanation types and formats to user needs and cognitive preferences, improving interpretability and usability. User feedback from site supervisors and engineers indicates a preference for visual dashboards and scenario-based prompts rather than technical model outputs. Such evidence shapes the explanation delivery by using counterfactual explanations in a visual format (e.g., 'If worker turnover were reduced by 30%, the risk level would shift from High to Moderate'), helping users understand and act on model outputs intuitively. ***XAI Techniques*** – CFEs (e.g., DiCE and Alibi-CF), and LIME, SHAP visualizations
- ***Validating Explanation Usefulness and Reliability***: Uses empirical user feedback and testing to assess whether explanations enhance understanding, trust, and decision accuracy. User testing reveals that site managers interpret SHAP-based explanations correctly in only 60% of cases, while simpler rule-based outputs (e.g., 'Risk increased due to design delays and poor subcontractor performance') improve comprehension and decision accuracy. This empirical evidence guides the refinement of the explanation mechanism to ensure the output is not only technically valid but also reliably understood and used by non-technical stakeholders. ***XAI Techniques*** – Human-in-loop evaluation frameworks, Explanation satisfaction metrics (e.g., fidelity, and simulatability)
- ***Ensuring Transparency, Justifiability, and Domain Alignment:*** Aligns system logic with industry norms, standards, and regulatory requirements, allowing traceability and accountability. Safety assessments linked to compliance data and inspection logs enable users to justify their risk assessments during regulatory audits and internal reviews. ***XAI Techniques*** – Rule-based models (e.g., decision trees), Traceable surrogate models, and post-hoc explanation audits.

Figure 2. A means-end framework for XAI-based design of DSSs.

Our framework aims to ensure that the norms and standards for good belief-forming practices (epistemic normativity) align with rational selection and the use of compelling evidence and methods (instrumental rationality) to achieve an organization's epistemic goals. By integrating these two dimensions, epistemic practices become both normatively justified and practically effective, giving end-users confidence in the outputs and explanations provided.

The aim of XAI is to provide instruments (means) that generate explanations which are understandable and meaningful to end-users, thereby fulfilling their epistemic needs (ends)[24]. For instance, in the case of LIME, the epistemic end is to explain an individual prediction within a local region of the data space, with the means being a simplified, interpretable surrogate model that approximates the behavior of the original AI model in that region [24]. Similarly, CFEs aim to clarify the input–output relationship of a model, using hypothetical 'what-if' statements as the explanatory means [24].

As illustrated in our framework (Figure 2), evidence plays a pivotal role in informing the design of such XAI instruments and the broader DSSs. It does so by guiding the selection of relevant features and model structure, shaping the form and modality of explanations, and validating their usefulness and reliability. In doing so, evidence ensures that explanations are not only technically accurate but also transparent, justifiable, and aligned with domain-specific causal reasoning, making them both trustworthy and actionable in practice.

Not only does a means–end epistemology underpin existing XAI instruments, but it also provides a valuable framework for the development of new tools, specifically by guiding critical design questions such as: *what* should be explained (subject), *to whom* it should be explained (stakeholder), *how* it should be explained (instrument), and *why* it should be explained (goal) [2], [6], [24], [26]. As previously discussed, different stakeholders, such as engineers who develop models, operators who implement them, and end-users who rely on their outputs, will inevitably have diverse explanatory needs, often requiring distinct explanation strategies or instruments to satisfy their respective epistemic goals [2], [6], [24], [66], [97], [98]. A detailed examination of these stakeholder-specific desiderata, particularly in the context of construction and infrastructure projects, is provided in Love *et al*. [6].

The suitability of XAI instruments depends on what should be explained and to whom the explanation is given. However, the suitability of an XAI instrument in providing an explanation can only be determined if the epistemic end being pursued by a stakeholder is specified in detail before acquiring evidence and constructing a DSS [24]. Sometimes, the same stakeholder may require different explanations, so they must be tailored to meet their specific needs. Consideration may also need to be given to *algorithmic recourse*[2], which can result in changes to the design of the DSS and a re-evaluation of evidence [99-101].

## 5.0 Evidence-Based XAI and DSS Design

Finding and using evidence to support XAI and the design and development of a DSS is crucial for providing trustworthy explanations about AI model outputs. The means-end framework in Figure 2 provides a pathway for end-users to realize their desired epistemic goals. End-users need to feel confident that their decisions are based on explanations derived from the best available evidence, which informs and helps them understand *why* a given output occurs. Likewise, this would also be the case for all other stakeholders.

[2] Algorithmic recourse refers to the ability to interpret algorithmic decisions enabling decision-makers to change outcomes within the context of CFE, for example.

### *5.1 Post-hoc Explainability*

At this juncture, the shortcomings of some prevailing XAI post-hoc explainability instruments need to be pointed out, as they can only provide end-users with *partial* explanations and do not provide answers to *why questions* [6], [26]. Post-hoc explainability (i.e., model-agnostic) instruments, such as LIME and SHAP, aim to provide an explanation of a 'black-box' algorithm using an interpretable predictor to make visible its internal workings by tracing its path-dependency (i.e., a given function(s) is related to an output). In doing so, an explanation is provided of how, rather than why, an output is generated [26]. Such a situation arises because post-hoc explainability is transparency-conditional, with explanations and predictions bound to a mediating predictor rather than the inner workings of the 'black box' [26]. As a result, an explanation is needed for *how* a formal representation between a 'black box' model and the interpretable predictor must exist. In this instance, a formal representation to provide structured ways to understand and quantify the relationship between complex and simplified models is required and created using concepts such as [102-105]:

- *Isomorphism* between a 'black box' model $f$ and an interpretable predictor $g$ implies a one-to-one correspondence where the structure and behavior of $f$ and $g$ are identical. For every input $x_1$, the output of $f$ matches the output of $g$ exactly, and their internal decision-making processes are structurally equivalent. So, let $\mathcal{X}$ be the input space, and $\mathcal{Y}$ be the output space. The models $f$: $\mathcal{X} \rightarrow \mathcal{Y}$ and $g$: $\mathcal{X} \rightarrow \mathcal{Y}$ are isomorphic if there exists a bijective function $h$: $\mathcal{X} \rightarrow \mathcal{X}$ such that:

  $$\forall x \in \mathcal{X}, f(x) = g\big(h(x)\big) \text{ and } g(x) = f\big(h^{-1}(x)\big) \qquad \text{[Eq.1]}$$

  If $f$ is a neural network and $g$ is a rule-based model that exactly replicates the decision boundaries and outputs of $f$ for all inputs, they are isomorphic.

- *Partial isomorphism* between a 'black box' model $f$ and an interpretable predictor $g$ indicates that the models are isomorphic only within a subset $\mathcal{S} \subseteq \mathcal{X}$ of the input space; that is, $g$ accurately mimics $f$ within $\mathcal{S}$. Thus, a formal representation is expressed as:

  $$\forall x \in \mathcal{S}, f(x) = g(h(x)) \text{ and } g(x) = f(h^{-1}(x)) \quad \text{[Eq.2]}$$

  If $f$ is a complex model used for a wide range of inputs but $g$ is a simpler model that accurately mimics $f$ for a specific type of input (e.g., a specific range or category), they exhibit partial isomorphism over that subset.

- *Similarity* between a 'black box' model $f$ and an interpretable predictor $g$ implies that $g$ approximates $f$ such that their outputs are close in a predefined sense (e.g., distance metric or statistical measure). This output does not require a one-to-one correspondence but rather a high correlation or similar statistical properties. So, let $\mathcal{X}$ be the input space, and $\mathcal{Y}$ be the output space. The models $f: \mathcal{X} \to \mathcal{Y}$ and $g: \mathcal{X} \to \mathcal{Y}$ are similar if there exists a similarity measure $d: \mathcal{Y}\, x\, \mathcal{Y} \to \mathbb{R}$ such that for some small $\epsilon \geq 0$:

  $$\mathbb{E}_{x \sim \mathcal{X}}[d(f(x), g(x))] \leq \epsilon \quad \text{[Eq.3]}$$

  Where $d$ could be any appropriate distance measure (e.g., Euclidean distance [107] and Kullback-Leibler divergence [108]). Suppose $f$ is a deep neural network and $g$ is a linear regression model trained to approximate $f$ outputs. Their predictions are similar if the average difference between their outputs is within an acceptable range $\epsilon$.

Putting aside the brief description above that illustrates *how* an explanation is generated, Hassija *et al*. [103] discuss the advantages and disadvantages of post-hoc XAI instruments, thereby providing readers with further insight into these issues.

To answer a *why* question requires moving beyond the navigation of an algorithmic path to explain the causal-mechanical or inferential dependence between X and Y and the corresponding effect [26], [105]. Accordingly, an established explanation needs to be supplemented with objective relations of dependence, such as a causal linkage between X and Y [26]. Therefore, multiple and complementary sources of evidence are essential to demonstrate causal linkages in this instance.

### *5.2 Hierarchy of Evidence*

The quality of evidence directly influences the reliability and validity of a generated explanation. Evidence hierarchies, such as that presented in Table 1, rank research and evaluation designs according to methodological rigor. As Table 1 shows, the greater the number of high-quality studies included in an analysis and the more rigorous their design, the higher the evidence rating. Systematic reviews, meta-analyses, and randomized controlled trials (RCTs) are typically placed at the top of such hierarchies in fields such as medicine.

While the hierarchy provides a useful starting point, its direct application to construction is limited by the sector's VUCA environment, where controlled trials and meta-analyses are rarely feasible. To address these methodological realities, we adapt the hierarchy to emphasize mechanism-rich longitudinal field studies, quasi-experimental interventions, and mixed-method causal analyses, forms of evidence most suited to the epistemic and practical constraints of construction (see Table 3). In doing so, the adapted hierarchy is aligned with our means–end framework, ensuring that evidence weighting reflects what is both achievable and epistemically valuable in construction research. Figure 3 illustrates how such evidence can be used within the means–end framework to inform the generation of a CFE.

Table 1. Hierarchy of evidence to support XAI design

| Level | Evidence Indication | Strength of Evidence | Description of Evidence |
|---|---|---|---|
| 1 | It is shown that… | **Strongest** | Meta-analyses and systematic reviews of (randomized controlled trials) or experimental studies |
| 2 | | | Single experimental study (randomized, controlled) with prospective real-world cases considered by practitioners in real-world settings |
| 3 | It is likely that… | | A single quasi-experimental study (e.g., nonrandomized, with concurrent or historical controls) involving prospective real-world cases considered by practitioners in real-world settings |
| 4 | | | A single experimental study (randomized, controlled) with retrospective real-world cases considered by real practitioners in simulated/laboratory settings. |
| 5 | | | A single quasi-experimental study (e.g., nonrandomized, with concurrent or historical controls) involving retrospective real-world (or simulated) cases considered by practitioners in simulated/laboratory settings |
| 6 | | | A single quasi-experimental study (e.g., nonrandomized, with concurrent or historical controls) involving simulated cases considered by human participants but not real practitioners in laboratory settings |
| 7 | There are signs that... | | A supervised machine learning train/test studies with external validation (multiple datasets in longitudinal or cross-sectional/multi-site settings) |
| 8 | | | Supervised machine learning train/test studies with internal validation |
| 9 | Experts believe that… | | Consensus opinions of authoritative bodies (e.g., nationally recognized guideline groups with robust peer review processes, notified bodies, and standardization organizations) |
| 10 | | **Weakest** | Opinions of recognized experts and case studies demonstrating examples of decision-making in similar conditions |

Adapted from Famiglini *et al*. [10]

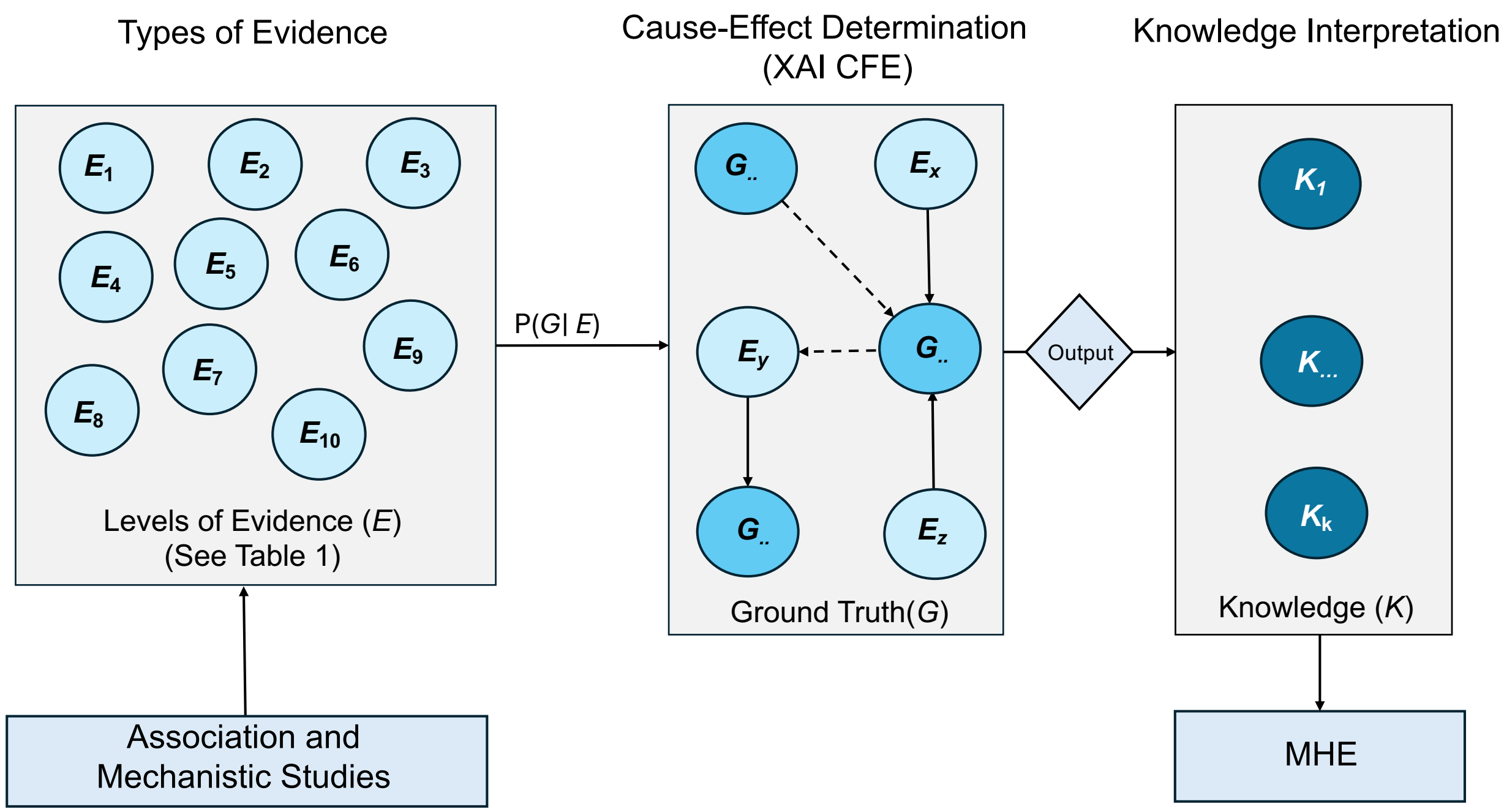

(a) Counterfactual instrument

The CFE mechanism consists of three components: (1) *Evidence identification* derived from a hierarchy of evidence, as noted in Table 1, and based on association and mechanism studies. Here, evidence is converted into information and weighted accordingly based on the varying explanation requirements of the end-user (stakeholders); (2) *Cause-effect determination* expresses how likely two events are to occur and predicts what will happen. The potential effects of changing causes are inferred; and (3) *Knowledge interpretation* aims to imagine 'what could happen', enabling counterfactual explanations to be examined. The generation of such answers may contribute to helping provide end-users (stakeholders) with suggested MHEs.

(b) Instrument description

Adapted from Shah and Choksuriwong [102]

Figure 3. Example of 'how' a CFE fits within a means-end framework

A hierarchy of evidence is essential for ensuring quality assurance (i.e., differentiating from high to low-quality evidence), transparency (i.e., well-defined criteria for evaluating and ranking evidence to ensure reproducibility), the handling of complexity (i.e., synthesizing multiple sources and assessing their relevance), and informed decision-making (i.e., drawing on reliable sources and prioritizing findings). As noted above, identifying the best available research evidence within the context of XAI in construction remains an open area for investigation. This is not to say that methods used to obtain the 'strongest evidence', for example, systematic reviews, meta-analyses, and RCT studies, are not performed in construction; quite the contrary. Fittingly, a wide range of areas, including occupational health and safety [109-116], structural optimization of concrete materials and structures [117-120], and building automation systems [121-123], have utilized such methods to support their respective lines of inquiry.

Hence, research methods and 'high-quality' evidence are readily available to support the design and development of XAI and DSSs in construction. Nonetheless, the evidence to support the design and development of interpretable DSSs in construction has been downplayed [6]. Equally, as Rudin [124] notes, this is a widespread problem with AI models in general.

The hierarchy of evidence in Table 1 provides a framework for interpreting the strength of evidence supporting a specific design choice, feature, or method for XAI [10]. It supports the generation of an interpretable DSS within an Evidence-Based Design (EBD) lens, focusing on [10], [125], [126]:

- *Research-driven decisions*: The systematic collection and analysis of empirical data and best practices derived from various study types (e.g., qualitative, quantitative, or a combination of both). However, as discussed below, multiple sources of evidence are required to arrive at an effective decision;

- *Epistemic outcomes:* As shown in Figure 2, the primary goal is to satisfy end-users' goals by incorporating their needs into the design of XAI instruments and the DSS. Key criteria that can be employed by end-users to evaluate outputs generated from their DSS include: (1) *accuracy* (i.e. reliable predictions or recommendations) and *precision* (i.e., minimal error rates); (2) *MHE*s provide clarity (i.e., easily understood), are contextually relevant (i.e., ensure scenario evaluation and expert review), answer 'why' and 'how' questions (i.e., provide explanations that actionable), and reveal their reasoning processes (i.e., seek feedback about transparency) [26]; (3) *response time* (i.e., the system should provide recommendations without excessive use of data and processing time); (4) *user experience and usability* (i.e., an intuitive interface that can be navigated with ease and minimal training where users having control how information is presented); (5) *scalability* (i.e. ability to handle larger datasets, higher user loads and complex tasks without a drop in performance); and (6) *compliance with regulations and ethical standards* (i.e., data privacy compliance and alignment with ethical norms to avoid harm, promote fairness, transparency and accountability); and
- *Continuous evaluation:* Assessing and reviewing the effectiveness of XAI instruments and the DSS in delivering interpretable and explainable outputs readily understood by end-users. The process of continuous evaluation can help identify problems and enable debugging and model improvements to be made (Figure 2). For example, questions such as 'Is additional information required to make outputs of an XAI instrument more understandable? and 'Does the DSSs design perform better than others in aiming to provide solutions to the same problem?' needs to be asked, and changes made to models, if required. Thus, the explicit aim of a continuous evaluation process is to ensure that the DSS is both efficient and effective, providing a user-centered solution.

The hierarchy of evidence presented in Table 1 serves as a 'heuristic' for making informed design choices to create XAI and DSS solutions that enhance decision-making effectiveness [10]. To reiterate, decisions should *not* be based on a single evidence source. Instead, several sources require consideration, as construction projects are procured in VUCA environments, where decision-making is subjected to constant change and the need to adapt to varying contexts [127]. Thus, access to different types of evidence that can demonstrate an association between variables and justify 'how' something occurs will enable the design of XAI instruments and DSSs that achieve their desired epistemic goals.

### *5.2 Evidential Pluralism*

With multiple sources of evidence needed to establish knowledge claims and explain outputs, the theory of *Evidential Pluralism*—an account of the epistemology of causation that maintains that to establish a causal claim, there needs to be an *existence* of a correlation and mechanism (Figure 4)—is utilized within the means-end framework to provide XAI instruments and the DSS with the means to ensure it is grounded on *explainable causal discovery* [28], [128], [129]. While Evidential Pluralism aligns with mixed methods, it is fundamentally different, as it provides a normative philosophical theory about what *needs* to be established to establish the presence of causation contributing to the propagation of MHEs for DSSs outputs [128]. Simply put, mixed methods are a methodological orientation employing variations in qualitative and quantitative data, methods, and designs [77].

Figure 4 presents the crux of Evidential Pluralism, indicating that to establish a causal claim, both a correlation and the existence of mechanisms must be demonstrated. A correlation and linking mechanism underpin the claim that *A* is a cause of *B* [128], [130]. Notably, association studies identify and analyze statistical associations between variables, helping to understand whether and how variables are related. They cannot prove causation; instead, they focus on correlation (i.e., the strength and direction of a relationship between variables) and tests of significance (i.e., determining whether a relationship in the data is statistically significant).

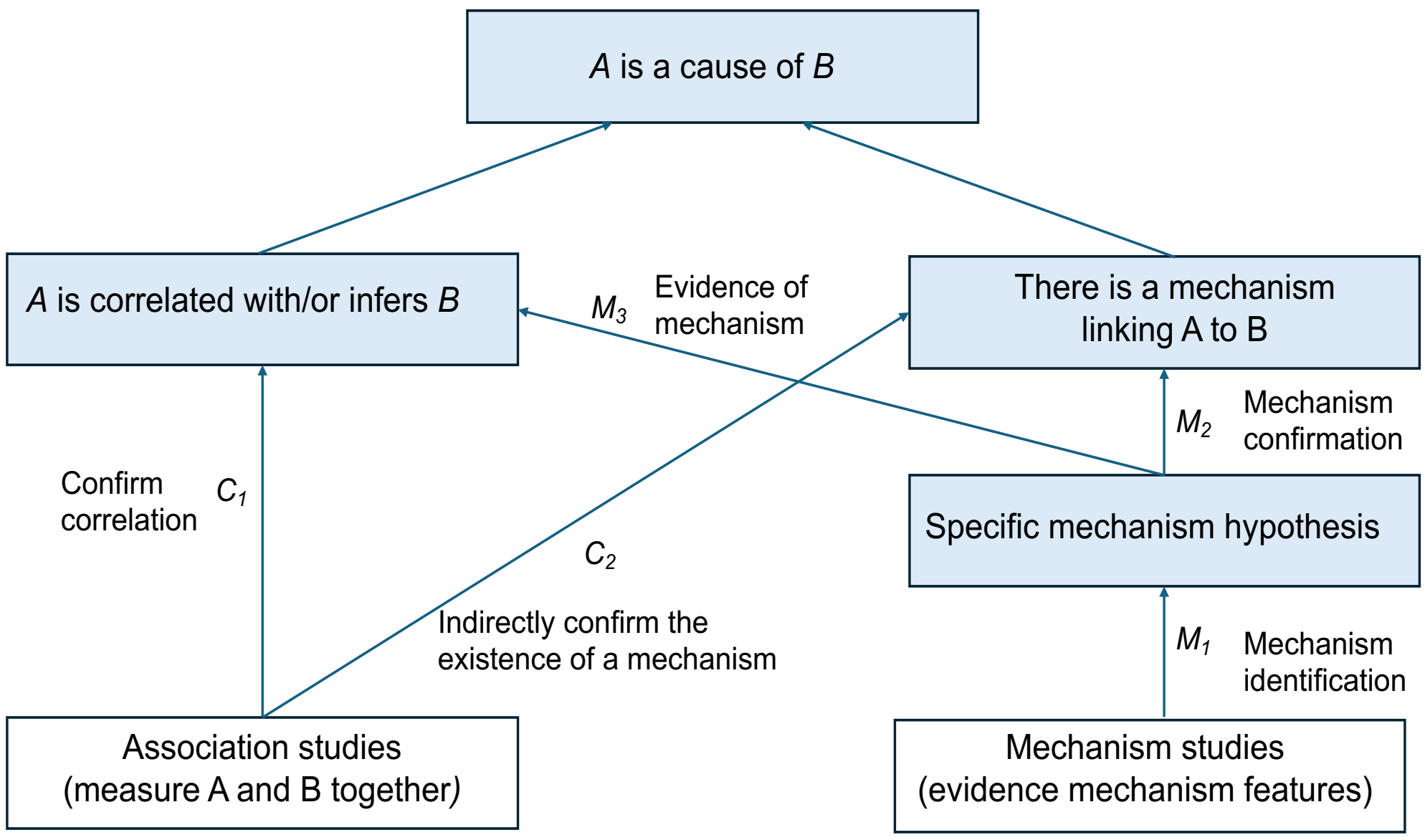

Adapted from Williamson [129: p.45]

Figure 4. Evidential pluralism: Assessing a causal claim

Each XAI method contributes distinct types of evidence: SHAP and LIME provide associative evidence on feature importance, CFEs offer quasi-mechanistic insight into causal levers, and surrogate models bridge local and global interpretability. Aligning these techniques with the evidence hierarchy ensures that their outputs are not only technically valid but epistemically robust, enabling users to act confidently on system recommendations.

### 5.2.1 Association Studies

Association studies can test a hypothesis ($C_1$) that a putative cause and effect are correlated with potential confounders (Figure 4). Similarly, they can be used to indirectly verify if a mechanism accounts for a correlation ($C_2$). Controlling for multicollinearity (e.g., removing highly correlated predictors or applying Principal Component Analysis to reduce dimensionality) and confounders (e.g., adding covariates to a regression model to adjust for their effects) is critical for ensuring the validity of a study, as it provides the ability to infer that there is a mechanism of action giving rise to a correlation.

With association studies, there is a need to be mindful and deal with confounding variables; caution also needs to be taken with 'tests of significance' (i.e., hypothesis testing) as they can influence the reliability of evidence when inferring a causal claim as: (1) $p$-values being misinterpreted: (2) thresholds of $p < 0.05$ (or greater), for example, for statistical significance resulting in binary thinking (significant *versus* non-significant) rather than a nuanced understanding of the data and the actualities of practice; and (3) a lack of reproducibility of results due $p$-values being sensitive to varying conditions. In addition to these limitations, which can impact the reliability of evidence used to design and develop XAI and DSSs, there is also a need to consider how well outputs correlate with expected outcomes or ground truths when association studies are relied upon in areas such as: [6], [129]:

- *Performance evaluation*: Precision, recall, and accuracy metrics, essentially quantifying the association between the predictions of an AI model and ground truth labels, have been the mainstay in construction for justifying a model's performance [131], [132]. However, there are limitations to these performance metrics, which can impact their accurate interpretation. For example, precision and accuracy metrics can be misinterpreted by imbalanced datasets and are insensitive to false negatives. Likewise, a trade-off between recall and precision often prevails. Hence, maximizing recall can result in a decrease in precision and *vice versa*.

- *Bias detection:* Examining whether a model's decisions are correlated with sensitive attributes such as demographics (e.g., race, gender, and groups) and individual characteristics (e.g., age and physical qualities) [133]. If an AI model's decisions disproportionately (dis)favor a particular individual or group, this can indicate a biased association that needs to be addressed. Several methods can be used to identify and address bias, depending on the context of construction (e.g., detecting unsafe behavior [133], posture analysis [134], and workforce activity [135]), ranging from statistical and predictive parity checks to more complex intersectional fairness metrics. Incompleteness of data, incorrectly if groups have similar statistical outcomes, intersectional complexity, an inability to cater to varying contexts in which a model operates (i.e., generating false positives or negatives), and changes in data are just a few of the limitations that must be addressed to overcome the detection of bias in DSSs; and
- *Causal inference:* Emphasis is based on determining a cause-and-effect relationship between variables and establishing a better understanding of the directional influence between variables than correlation. Causal inference models commonly used in XAI studies in construction are Directed Acyclic Graphs (DAGs) and variants thereof [39], [64], CFEs [40], and Granger Causality [136]. However, models of this type are subject to several drawbacks, as confounders can go unmeasured. Multiple explanations can also fit the observed data, introducing a degree of ambiguity and hindering the ability to generalize causal relations, particularly in high-dimensional datasets. Aside from these disadvantages, causal inference models *cannot* demonstrate causality in an absolute sense, reinforcing the need to identify specific linking mechanisms to verify a claim.

While association studies take a coarse approach to establishing a link between cause and effect, mechanistic studies can complement them by uncovering *how* the instances of *A* are responsible for those of *B*.

5.2.2 Mechanism Studies

Mechanism studies are essential for informing the design of XAI instruments and the resultant explanations, as well as assessing AI models, as they provide a detailed understanding of the internal processes and pathways that drive their predictions. Within the context of Evidential Pluralism, as noted in Figure 4, mechanism studies do not measure *A* and *B* together. Instead, they seek to unearth the links or components of the complex mechanism. As a result, they can provide a nuanced perspective on putative causal relationships as they examine the specific processes, structures, and systems that explain *how* an event occurs and how an AI model works, supporting the generation of MHEs [128].

The existence of a mechanism is confirmed, as noted in Figure 4, by employing hypotheses that posit features likely to link *A* and *B* ($M_2$). According to Williamson [129], a *specific mechanism hypothesis* should articulate the components of the mechanism, the interactions between them, and the expected outcomes under certain conditions. Still, it does not necessarily need to identify every feature. Additionally, features can be identified and confirmed by previous mechanistic studies ($M_1$), such as observational evidence (Table 1), and used to support correlated variables ($M_3$).

Identifying mechanisms helps XAI simplify complex models into more interpretable components or surrogate models (e.g., decision trees, rule-based models, and Generalized Additive Models), which not only help explain predictions. However, it can also be used for tuning parameters and architectures, as well as debugging, thereby contributing to improved performance. Specific areas where mechanism studies can be used to assess the performance of XAI include:

- *Explanation accuracy and completeness*: Assessing how accurately instruments attribute contributions to different features. Methods such as LIME and SHAP can be used to ensure that XAI instruments correctly represent the importance of each feature. Indeed, LIME and SHAP are prevalent XAI approaches in construction [6], [137], [138], but they come with their own set of challenges. While LIME approximates an AI model locally around a specific instance using a straightforward, interpretable approach (e.g., a linear model), it may not always accurately reflect its global behavior, especially if its decision boundaries are non-linear. Moreover, LIME relies on generating perturbed samples around the instance being explained. The stability of the explanations can be affected by the choice of sampling strategy, which might lead to variability in explanations across different runs or settings; this point was also identified above in Section 3. In the case of SHAP, there is an assumption of a linear relationship between its values (i.e., calculated by comparing a model's prediction with and without a particular feature present) and a model's predictions; however, this assumption does not hold for models and thus should be treated with caution when interpreting them. It should also be noted that while SHAP values attribute feature importance, they *do not* provide causal insights. As a result, understanding how they causally interact will invariably be required using techniques such as DAGs, CFEs, or Structural Equation Models [139]. In this instance, the need for both association and mechanism-based studies is explicit, *writ large*.
- *Model transparency and end-user understanding*: The consistency of explanations provided by XAI instruments to end-users is central to ensuring model transparency [140]. Here, consistency refers to the homogeneity of explanations obtained from a model when applied to similar inputs and the varying contexts of user needs and expertise [140]. The consistency of explanations can be evaluated, depending on a model's type and purpose, using an array of techniques such as perturbation analysis, feature removal, K-NN, correlational analysis, similarity metrics (e.g., in documents using cosine similarity, Euclidean distance, or Jaccard index [141]), saliency maps, and partial dependence plots. Combining these approaches can enhance the reliability of explanations and aid in assessing whether they are reasonable, plausible, and aligned with epistemic goals.

In sum, mechanism studies are crucial for evaluating XAI and DSSs, as they offer a deeper understanding of the internal workings of models, which is necessary to support association studies in generating explanations of outcomes for end-users. When LIME and SHAP are applied within an evidence-based framework, their outputs should be interpreted with caution. Although they reveal associations between input features and model predictions, they do not in themselves confirm underlying mechanisms. Their stochastic nature means that explanations can vary depending on sampling strategy or data perturbation. Accordingly, they are most useful when triangulated with other sources of evidence, such as causal modeling or process-tracing interviews, to ensure that the resulting explanations are robust and mechanistically meaningful. Within the adapted evidence hierarchy (Table 3), such outputs correspond to associative evidence and should be weighted accordingly when informing DSS design and explanation generation.

## 6.0 Discussion

The means-end framework, identified in Figure 2, provides a basis for identifying and structuring evidence to create *bona fide* MHEs, which can provide end-users with confidence that the outputs generated from DSSs are reliable. Consequently, this will enable construction organizations to realize the benefits and business value of XAI and DSSs [2], [142].

### *6.1 Developing Meaningful Explanations*

While a means-end lens provides a new conceptualization for incorporating evidence into the design of a DSS, the quality of the data and its preparation (e.g., data cleaning, feature engineering, transformations, and splitting), as well as the association and mechanism studies, will influence the effectiveness of the MHEs generated. Thus, researchers in construction will need to give credence to the *strength, value,* and *utility* of evidence that has been captured and crafted to provide MHEs [26]. The strength, value, and utility of evidence required will vary with its availability and the context of a DSS application in construction, for example, whether it is being used to manage cost and schedule performance, assess the risks associated with ground settlement during tunnel boring machine operations, inform retrofit decisions, or predict concrete strength. An MHE should comprise three components [2]:

1. *Explanatory power:* How well a model can address different dimensions of explanation in varying contexts [143] and by answering a series of CFEs (e.g., what-if-things-had-been-different questions) that illustrate the theoretical and pragmatic relevance of an explanation [144];
2. *Control the process of explaining*: People play a critical role in determining whether explanations adhere to epistemic and normative standards. The design of DSSs is not value-free, as it involves stakeholders' varying economic, ethical, legal, political, and social issues, which must be considered, given that their beliefs, values, and expectations will differ. Construction organizations should always be prepared to question the explanations provided by DSSs. Expanding on this point within the context of AI algorithms used to support robots in operations, Santoni de Sio and van den Hove [145] argue that humans should be responsible for and control all decisions made. Thus, to maintain control over decisions, construction organizations need to be able to specify the conditions under which an MHE satisfies its users of a DSS. Identifying the properties of an explanation relevant to the end-users' epistemic goals so that they can be accordingly assessed; and

3. *Evaluating the suitability of evidence to inform XAI*: Integrating evidence with explanatory power will not only enable reliable assessments of new knowledge but also foster learning, understanding, and conceptual coherence, allowing end-users to evaluate various MHEs for different situations that may arise during construction.

If end-users in construction cannot make sense of MHEs by drawing on and combining them with prior knowledge, then what purpose do they serve? [26] Although there is a need to incorporate XAI into AI models in construction, MHE must be capable of providing an understanding based on evidence and a well-defined structure.

### *6.2 Evaluating XAI in Construction*

The integration of XAI into DSSs holds promise for enhancing decision quality, increasing transparency, and fostering user trust. However, to ensure these benefits are realized, XAI systems must be rigorously and systematically evaluated [2]. Evaluation must extend beyond technical performance and address how well explanations serve the practical, cognitive, and normative needs of users within the VUCA environment of construction. We identify the various types of XAI evaluation methods and applications in construction in Table 2 and briefly discuss them below

### 6.2.1 Cognitive Validity: Ensuring Human Comprehension

Evaluation is critical to ensure that explanations are cognitively accessible to users such as construction managers, site and project engineers, and project and safety inspectors. As construction is inherently subject to production pressure, users of DSSs often do not have the time or capacity to decipher overly complex or technical explanations. User-centred evaluation methods, including surveys, think-aloud protocols, and cognitive load assessments, can help determine whether explanations are easily understood, actionable, and presented at the appropriate level of detail. Without this, XAI may hinder rather than help decision-making.

### 6.2.2 Epistemic Alignment: Supporting Meaningful Understanding

In construction, where causal relationships are often uncertain and context-dependent, explanations must do more than highlight model features, they must help users understand why or how an outcome occurred [2]. Evaluating epistemic soundness ensures that explanations align with domain knowledge and established causal reasoning, rather than presenting shallow statistical correlations [28]. This alignment enhances users' ability to interpret outcomes, diagnose problems, and justify decisions to stakeholders, especially in tasks such as delay attribution or safety and quality analysis [133].

### 6.2.3 Robustness and Reliability: Promoting Trustworthy Use

Trust in AI systems is undermined if explanations change unpredictably or fail under slightly different conditions. Evaluation of explanation stability, consistency, and fidelity (i.e., how faithfully the explanation reflects the model's true reasoning) is crucial for identifying unreliable behavior. In construction applications where inputs may be noisy, incomplete, or context-sensitive, such robustness checks are essential to ensure the dependability of AI support in real-world scenarios [77], [79].

Table 2. Methods to evaluate XAI

| Evaluation Method | Technique | Description | Importance | Application in Construction |
|---|---|---|---|---|
| **User-Centred** | *User Studies / Surveys* | Gather feedback from end-users (e.g., engineers, project managers) to assess their understanding, trust, and satisfaction with the explanations. | Ensures that explanations meet the cognitive and informational needs of real users in construction environments. If users cannot understand or trust an explanation, it undermines the utility of XAI in practice. | Evaluating whether site managers trust cost overrun or safety risk predictions |
| | *Think-Aloud Protocols* | Users verbalize their thought process as they interact with an XAI system. | Reveals how users interpret explanations in real time, exposing potential mismatches between model logic and human reasoning. This helps refine explanations to align better with user expectations and workflows. | Ensuring the interpretability of risk models under time-constrained decisions |
| | *Cognitive Load Assessment* | Measures the mental effort required for users to understand and act on an explanation. | In high-stress construction contexts, explanations must be cognitively efficient. High cognitive load can lead to misinterpretation or non-use of XAI outputs, reducing its effectiveness. | Ensuring the interpretability of risk models under time-constrained decisions |
| **Ground Truth Comparison** | *Expert Comparison* | Compares model-generated explanations to those provided by domain experts. | Validates whether the explanation is plausible and grounded in the construction domain logic. Helps prevent black-box outputs that conflict with known causal relationships or best practices. | Verifying delay or rework explanations *via* expert panels |
| | *Explanation Consistency* | Tests whether small changes to inputs produce stable and consistent explanations | Construction environments are sensitive to minor contextual variations. Unstable explanations may erode user confidence, particularly in safety-critical or cost-sensitive applications | Checking SHAP-based reasoning to predict quality |
| **Quantitative Metrics** | *Fidelity / Faithfulness* | Measures how accurately the explanation reflects the model's internal decision-making. | Prevents misleading users by ensuring the explanation is a truthful representation of how the model actually works. This is essential for accountability and auditability in construction DSS. | SHAP explanations for model outputs in defect classification |
| | *Stability / Robustness* | Assesses how explanations respond to small perturbations in the input data. | Critical in construction applications where noisy, incomplete, or uncertain data are common. Robust explanations are more reliable and less prone to manipulation. | Evaluating risk models for safety assessments |
| | *Sufficiency / Comprehensiveness* | Evaluates whether the explanation includes the minimum or complete set | Ensures explanations are neither too vague nor overly complex, providing the right level of detail | Testing which features matter most in rework prediction |
| | | of important features needed to justify the output. | for effective human understanding and action. | |
| **Task-based Evaluation** | *Scenario Testing* | Compares user task performance with and without explanations in realistic or simulated tasks. | Measures the practical utility of XAI in supporting better decision-making under realistic construction conditions (e.g., project planning, resource allocation). | Scheduling or resource allocation using explainable models |
| | *Decision Impact Analysis* | Evaluates how the explanation influences user choices or judgments. | XAI should not just explain but improve decisions. This method tests whether explanations lead to more informed, consistent, or optimal decisions—especially critical in risk management and quality assurance. | Assessing change in supplier or subcontractor selection after exposure to explanations |
| **Epistemic Evaluation** | *Epistemic Soundness* | Assesses whether the explanation aligns with scientific, causal, or domain-specific reasoning. | In construction, where causal factors (e.g., for rework or delays) are complex and multi-layered, epistemically sound explanations support better diagnosis and learning—not just prediction. | Explaining delay causes in a causally meaningful way |
| | *Normative Justifiability* | Determines whether the explanation meets ethical and rational standards for justification | Promotes transparency, fairness, and responsibility in AI-supported decisions, especially when they affect worker safety, contractual obligations, or compliance. | Ethical justification in AI-based claims or compliance models |

#### 6.2.4 Decision Impact: Measuring Practical Utility

Explanations should not only clarify outputs but also support better decision-making. Task-based evaluation methods, such as scenario testing and decision impact analysis, can be used to measure how explanations affect user behavior, performance, and outcomes. For instance, if an explanation of risk prediction helps a planner reallocate resources more effectively, this demonstrates that the explanation is not only understood but also useful; these forms of evaluation ground XAI design in measurable improvements in user performance and project results.

#### 6.2.5 Normative Justifiability: Upholding Ethical and Legal Standards

Decisions in construction often have ethical, legal, and contractual implications. Evaluating normative justifiability ensures that explanations meet standards for fairness, transparency, and accountability. Such considerations are particularly important in contexts such as compliance audits, safety investigations, or dispute resolution, where the rationale behind a decision may be subject to external scrutiny. Explanations must therefore be not only technically accurate but also justifiable in ways that stand up to institutional or regulatory review.

Considered collectively, these dimensions of evaluation, cognitive validity, epistemic alignment, robustness, decision impact, and normative justifiability form a comprehensive foundation for designing and deploying effective XAI systems in construction. Without rigorous evaluation, there is a risk of implementing systems that may appear sophisticated but fail to support users in meaningful or trustworthy ways. Hence, evaluation must be treated as a core pillar of XAI design, enabling systems that are not only explainable in theory but also explanatory, usable, and justifiable in practice. The operationalization of MHEs can be achieved through empirical indicators such as explanation satisfaction scores, cognitive load measures, and task-performance improvements. These metrics provide a structured basis for future empirical studies to determine whether an explanation is genuinely meaningful, supporting user understanding, decision quality, and trust, rather than merely technically accurate.

### *6.3 Illustrative Application*

To translate our conceptual means–end framework into a practical pathway for researchers and practitioners, we offer an illustrative, step-by-step application demonstrating how evidence can be integrated into the design, explanation, and evaluation of an XAI-enabled DSS. As a starting point, the adapted evidence hierarchy, presented in Table 3, provides a structured basis for weighing association and mechanism evidence, ensuring that higher-quality, contextually relevant studies are prioritized when informing model design. The suggested five-step process is as follows:

1. *Evidence Identification and Weighting*: For a DSS predicting schedule delays in tunneling projects, evidence is gathered from two primary sources: (a) association studies (e.g., historical project data showing correlations between delays, subcontractor reliability, and design finalization dates) and (b) mechanism studies (e.g., process-tracing interviews revealing how late design approvals propagate through procurement and scheduling, amplifying delay risk). The adapted evidence hierarchy presented in Table 3 is then applied to weight these sources, giving priority to high-quality longitudinal field data and mechanistic studies instead of single-case anecdotal evidence.
2. *Model Design and Feature Selection*: Evidence directly informs feature engineering: variables with strong support (e.g., geotechnical complexity and subcontractor reliability) are retained, while those with weak or spurious associations are excluded. As a result, causal interpretability is improved, and model noise is reduced.

3. *XAI Technique Selection and Mapping*: XAI instruments are selected to align with epistemic goals and the type of evidence required. LIME and SHAP provide associative evidence by estimating feature contributions to individual predictions, while CFEs contribute mechanism-oriented insight by revealing how changes to inputs causally affect outputs. Surrogate models, such as decision trees, can bridge association and mechanism by approximating global decision logic. Each technique is mapped to the evidence category identified in the adapted hierarchy (Table 3) and to the epistemic need it satisfies, whether supporting understanding, trust-building, or actionability.
4. *Explanation Delivery*: Explanations are communicated using dashboards with visual CFEs (sliders or scenario trees) and natural language summaries tailored to stakeholder needs (e.g., actionable insights for site supervisors, aggregated insights for project managers).
5. *Evaluation of MHEs*: Explanations are iteratively evaluated against the five dimensions identified earlier: cognitive validity, epistemic alignment, robustness, decision impact, and normative justifiability. Methods include think-aloud protocols, explanation satisfaction scales, stability testing, and decision impact analysis, ensuring that MHEs are not merely normative ideals but empirically testable outcomes.

Each step of this illustrative process corresponds to the core components of the means–end framework identified in Figure 2, demonstrating how epistemic means (evidence, modeling, explanation delivery) support epistemic ends (understanding, trust, and decision impact). The five-step process thus provides a clear, actionable pathway for operationalizing the framework and translating its conceptual insights into practice. Recognizing these challenges establishes a foundation for the limitations outlined in the following section and helps identify priorities for future empirical validation.

Table 3. Adapting the evidence of hierarchy for construction

| Level | Adapted Evidence for Construction | Role in DDSs/XAI Design |
|---|---|---|
| 1 | Systematic reviews/meta-analyses of quasi-experiments or longitudinal field studies | Meta-analysis of rework causes across projects to inform model feature selection and design choices |
| 2 | Quasi-experiments in live projects (A/B process interventions) | A/B testing of safety interventions; provides strong causal inference when randomization is impractical |
| 3 | Mixed-method longitudinal case studies combining association and mechanism evidence | Multi-year studies integrating interviews, process tracing, and statistical modeling to capture causal pathways |
| 4 | Observational field studies with causal modeling (e.g., DAGs, SEMs) | DAGs identifying root causes of tunneling delays; supports feature relevance |
| 5 | Cross-sectional surveys triangulated with expert panels | Survey + Delphi panel validation to provide indicative evidence for early-stage model design |
| 6 | Validated laboratory simulations | Virtual construction simulations testing decision-support interventions (limited external validity) |
| 7 | ML model performance with external validation | Risk prediction model validated on an independent project dataset; provides associative evidence |
| 8 | ML model performance with internal validation | Exploratory model development: lowest reliability, suitable for hypothesis generation |
| 9 | Consensus statements by recognized industry bodies | National or sectoral guidelines; aligns model assumptions with industry norms |
| 10 | Individual expert opinion or case vignettes | Expert judgment on contingency sizing; used for hypothesis generation only |

Applying the adapted hierarchy ensures that the evidence used to design, explain, and evaluate DSSs is both context-appropriate and epistemically robust. By prioritizing mechanism-rich longitudinal studies, quasi-experimental interventions, and triangulated causal analyses, DSS and XAI system designers can anchor model features and explanations in evidence that reflects the realities of practice in construction rather than relying solely on statistical associations. In doing so, they strengthen the credibility of model outputs, actively support the generation of MHEs, and empower practitioners to understand, trust, and act on DSS recommendations with confidence.

### *6.4 Availability of Data for Model Development*

It can be observed within the mainstream CEM literature that the number of papers focusing on AI has increased exponentially over the last decade [1], [6], [146], [147]. When new deep learning AI algorithms are developed (e.g., the You Only Look Once series [148]), a flurry of papers are published in the CEM-related literature claiming to provide accurate and reliable results for a given problem. Nonetheless, works of this ilk often offer no evidence to underpin their rationale and explain the results presented [149-152]. Furthermore, they are frequently based on experiments and small datasets, akin to a 'proof of concept', providing no insight into the realities of practice in construction and the limitations of the research [131].

High-quality and large datasets are essential for building effective DSSs that utilize deep learning algorithms [153]. While algorithmic advances are being made to improve prediction accuracy without using more data, the size used for training remains the most critical aspect of an AI model's development [154], [155]. Collecting large datasets is costly and time-consuming, particularly from construction sites, as they are typically unstructured, noisy, incomplete, and in varying formats [4]. Thus, to make headway in developing DSSs that can be effectively used by construction organizations and improve their decision-making, researchers must pool their resources and share data. Collaboration between researchers and organizations is necessary to establish a repository of shared evidence and datasets. Unfortunately, such collaboration appears to be missing within the CEM community.

Setting aside the need for collaboration, journals and their editors play a pivotal role in ensuring that the data used in AI studies are made available and, if sensitive, in a format that prevents identification of any organization or individual, thereby enabling replication and building upon the authors' claims. A condition of *Nature*, for example, is that authors must make materials, data, code, and associated protocols promptly available to readers without undue qualifications [155].

However, authors publishing studies on AI in construction rarely make their materials, data, code, and associated protocols accessible for review, instead opting to make them available upon request, a provision offered by journals. When authors requested access to AI training data to reduce bias and improve the generalizability of the models being developed, the corresponding authors of papers repeatedly ignored these requests. If individual privacy and intellectual concerns were associated with sharing the data, it is understandable why it could not be provided. Explicitly, this was not the case for the requested data.

Access to additional data to assist with training a model helps understand how it has learned and performs in real-world situations, enhancing the ability to develop explanations using XAI instruments and thus giving construction organizations the confidence to utilize DSSs in their operations. For this reason, it is strongly recommended that editors of mainstream CEM journals receiving AI-based papers request authors provide evidence (e.g., code and datasets) - perhaps a similar policy to *Nature* - to support the development of their models. By doing so, editors can make a valuable contribution to promoting much-needed collaboration, transparency, reproducibility, and ethical practices in the field, ultimately leading to advancements in research and technology.

### *6.5 Limitations*

Despite offering valuable theoretical insights, our review is subject to several limitations inherent in its methodological and conceptual approach. Using a narrative review allows interpretive depth and flexibility but lacks a systematic search protocol, which may introduce selection bias and limit the transparency and reproducibility of the review process. Prioritizing theoretical integration over empirical synthesis also constrains the generalizability of the findings, as the conceptual arguments are not directly validated through user data or case evidence.

Our means–end framework, while conceptually robust, remains abstract and may be challenging to operationalize without additional empirical grounding or practical guidance. The disciplinary focus on construction may further restrict the transferability of insights to other sectors with different epistemic or organizational demands. In addition, the analysis assumes a level of user agency and capacity to engage critically with AI systems that may not reflect the realities of professional practice, where time, resources, and technical expertise are often limited.

Beyond these methodological constraints, the study makes several important conceptual contributions that advance understanding of how XAI systems can be designed to meet the epistemic needs of end-users. It introduces the notion of MHEs, shifting the focus from purely technical explanations to those that are actionable, context-sensitive, and aligned with practitioners' decision needs, explaining why a decision was made, how outputs relate to inputs, and when recommendations can be trusted. By emphasizing the integration of multiple forms of evidence (e.g., mechanistic, statistical, and experiential) into XAI design, we highlight the underexplored role of theory in shaping XAI development for construction. In doing so, it positions end-users as active participants who can articulate their knowledge needs and demand transparency and accountability from AI systems. The means–end framework provides a structured lens for aligning AI explanation design with user goals and decision contexts, laying the groundwork for more user-centric and trustworthy AI in construction and other safety-critical industries.

A further limitation is that our means–end framework is presented conceptually and illustrated through hypothetical scenarios rather than through an empirical case study. This design choice is deliberate and consistent with the paper's aim as a narrative review and theory-building exercise. A worked, data-driven demonstration would require access to project records, stakeholder engagement, and iterative model development, activities beyond the scope of our paper. Future research will focus on operationalizing our framework in real construction projects, combining association and mechanism studies to generate CFEs and empirically evaluating MHEs using user-centred metrics such as cognitive load, explanation satisfaction, and decision impact. Such work will provide a rigorous validation of the framework's applicability and value in practice.

### *6.6 Implications for Future Research*

Future research is needed to develop further and empirically test the proposed means-end framework's ability to generate satisfactory explanations based on evidence. More specifically, the means-end framework must be populated with use cases from applications applied on-site (*domain*) for various areas such as defect detection using sensors, analysis of safety behaviors, predicting risks and costs, managing schedule performance, and checking adherence to codes and standards (*AI task*). Here, evidence from cases is used to inform and establish explanations of AI decisions, thereby improving trust and ensuring regulatory compliance (*XAI role*). In cases where potential defects or quality issues are detected, such as during tunnel construction, sensor data and evidence derived from historical data can be used to interpret AI-generated outputs using SHAP and LIME (*XAI techniques*). Here, SHAP can identify the key factors that contribute to a defect (e.g., human error, poor materials, and unfavorable environmental conditions). Likewise, LIME, in conjunction with evidence, can be used to explain how and why a particular section of a project (e.g., piling) experienced a defect, offering actionable insights into root causes.

Building a repository of use cases, which should be made openly available to researchers and others, will contribute to developing a generic, step-by-step implementation framework that can ensure the development of practical MHEs and support end-users' epistemic goals. Similarly, research must consider how the means-end framework can cater to the varying desiderata of stakeholders and produce MHEs. However, a significant challenge will be to acquire multiple sources of evidence to support AI tasks and the generation of MHEs, which will be relied upon to inform decision-making.

## 7.0 Conclusion

Using a narrative review, our paper addresses the research question: *How can evidence effectively inform the development of XAI systems and DSSs to ensure the epistemic needs of end-users are met?* Three sub-questions guided our inquiry. First, we examine *the role of evidence in XAI system design*, highlighting how different types of evidence, such as those from association and mechanism studies, support the reliability and interpretability of AI outputs, and advocate for a hierarchy of evidence to guide system development. Second, we examine *how XAI can deliver explanations that meet users' epistemic needs*, introducing the concept of MHEs, which includes understanding the rationale behind decisions, the influence of input changes on outputs, and whether users can trust and act upon the system's recommendations. Finally, we address *how construction organizations can engage with XAI despite not being system developers* by demonstrating how end-users can understand AI-driven decisions, articulate epistemic requirements to vendors and researchers, and critically assess DSSs to ensure alignment with their practical needs.

Given that the CEM literature has yet to engage substantively with the integration of evidence in the design of XAI instruments and DSSs, this study draws on contemporary research from other fields to inform its analysis. As construction organizations increasingly adopt both bespoke and commercial off-the-shelf DSSs, there is a growing need to embed XAI into their design to enhance transparency, fairness, and trust in system outputs. While the relevance of XAI is gradually being acknowledged within the CEM domain, the design of such systems must also incorporate robust evidence, derived from both association and mechanism-based studies, to support meaningful interpretation and informed decision-making.

To address this need, the paper proposes a means–end theoretical framework for integrating evidence into the development of XAI systems and DSSs. This framework is underpinned by Evidential Pluralism and structured around a hierarchy of evidence, thereby supporting explanation through causal discovery. It provides a systematic approach for aligning the design of DSSs not only with the epistemic needs of end-users but also with broader stakeholder goals.

From a practical standpoint, most construction organizations are not in a position to develop AI or XAI-based systems themselves, as their core focus remains on project delivery. While a few large firms, such as Obayashi Corporation, have in-house research and development capabilities, most remain technology users rather than creators. Nonetheless, the proposed framework equips organizations with a clear lens for understanding DSS design and ensuring that epistemic goals (i.e., clarifying recommendations, exploring variable interdependencies, and fostering trust in outputs) are effectively embedded in MHEs.

To fully leverage the benefits of XAI and DSSs, construction organizations must move beyond superficial rationales for AI-generated decisions and instead seek deeper explanations of why and how these outputs are produced. The means–end framework we have proposed in this paper not only provides a foundation for developing user-centric and functionally oriented AI systems in construction but also serves as a transferable model for other engineering domains seeking to integrate evidence into the design of explainable DSSs. By clarifying how evidence can be embedded in XAI design and proposing a structured means–end framework, our paper establishes the conceptual groundwork for future empirical testing, user studies, and implementation research across construction and other safety-critical sectors. Building on this conceptual foundation, the next critical step is to validate the framework in projects, integrating real-world data and engaging stakeholders to assess whether it reliably generates MHEs that improve decision quality. Such empirical work will ensure that XAI-enabled DSSs are not only theoretically robust but also demonstrably valuable in supporting understanding, trust, and actionable insights in practice.